\documentclass[12pt,preprint]{aastex6}

\def\apj{{ApJ}}
\def\mnras{{ MNRAS}}

\def\be{\begin{equation}}
\def\ee{\end{equation}}
\def\bea{\begin{eqnarray}}
\def\eea{\end{eqnarray}}

\usepackage{graphicx}

\def\gcn{{ GCN Circ}}

\begin{document}

\title{Evaluating the bulk Lorentz factors of outflow material: lessons learned from the extremely-energetic outburst GRB 160625B}

\author{Yuan-Zhu Wang$^{1,2}$, Hao Wang$^{1,2}$, Shuai Zhang$^{1,2}$, Yun-Feng Liang$^{1,2}$, Zhi-Ping Jin$^{1}$, Hao-Ning He$^{1}$, Neng-Hui Liao$^{1}$, Yi-Zhong Fan$^{1}$, and Da-Ming Wei$^{1}$}
\affil{
$^1$ {Key Laboratory of dark Matter and Space Astronomy, Purple Mountain Observatory, Chinese Academy of Science, Nanjing, 210008, China.}\\
$^2$ {University of Chinese Academy of Sciences, Yuquan Road 19, Beijing, 100049, China.}\\
}
\email{Email: liangyf@pmo.ac.cn(YFL), jin@pmo.ac.cn(ZPJ) and dmwei@pmo.ac.cn(DMW)}

\begin{abstract}
GRB 160625B is an extremely-bright outburst with well-monitored afterglow emission. The geometry-corrected energy is high up to $\sim 5.2\times10^{52}$ erg or even $\sim 8\times 10^{52}$ erg, rendering it the {\it most  energetic GRB prompt emission recorded so far.} We analyzed the time-resolved spectra of the prompt emission and found that in some intervals
there were likely thermal-radiation components and the high energy emission were characterized by significant cutoff. The bulk Lorentz factors  of the outflow material
are estimated accordingly. We found out that the Lorentz factors derived in the thermal-radiation model are consistent with the luminosity-Lorentz factor correlation found in other bursts as well as in GRB 090902B for the time-resolved thermal-radiation components. While the spectral cutoff model yields much lower Lorentz factors that are in tension with the constraints set by the electron pair Compoton scattering process. We then suggest that these spectral cutoffs are more likely related to the particle acceleration process and that one should be careful in estimating the Lorentz factors if the spectrum cuts at a rather low energy (e.g., $\sim$ tens MeV). The nature of the central engine has also been discussed and a stellar-mass black hole is favored.
\end{abstract}

\keywords{Gamma-ray burst, Lorentz factor, thermal, cutoff}

\section{Introduction} \label{sec:intro}

The outflows of Gamma-ray Bursts (GRBs) are generally considered to move relativistically to solve the ¡°compactness problem¡± \citep{1999PhR...314..575P,Kumar2015}. However, the Lorentz factor of the outflow is not an observable quantity. Several methods have been proposed to estimate the Lorentz factor ($\Gamma$) base on different hypothesis or fireball models: a lower limit can be obtained by requiring the Lorentz factor is large enough to make the observed most energetic photon not to annihilate \citep{1991ApJ...373..277K,1993A&AS...97...59F,1995ApJ...453..583W,1997ApJ...491..663B}. If cutoffs are observed on the high end of the spectra of prompt emissions, the exact values of Lorentz factor rather than lower limits can be derived by assuming the optical depth equals unity for photons with cutoff energies \citep{2001ApJ...555..540L,2011ApJ...729..114A,2015ApJ...806..194T}. Thermal components that accompany the underlying nonthermal emissions in several GRBs are thought to originate from the photosphere of fireballs; thus, they can also be used to determine the Lorentz factors \citep{2007ApJ...664L...1P,Ryde2010,Fan2011,Zou2015}. Note that these approaches are valid for the time-resolved outflow material as long as the spectra can be reliably measured.  Another kind of methods is to model the multi-wavelength afterglow based on the dynamics of fireball. In the thin shell case, the reverse shock is weak and the optical/X-ray emission is dominated by the forward shock emission with an almost constant Lorentz factor. Hence the peak of the optical/X-ray emission marks the deceleration of the fireball and can probe the Lorentz factor robustly \citep{1993ApJ...405..278M,Molinari2007,Jin2007,Xue2009,2010ApJ...725.2209L,Liang2015}. In the thick shell case, the reverse shock is strong and the Lorentz factor can be determined by a self-consistent modeling of the optical flash as well as the later afterglow emission \citep{1999A&AS..138..537S,2000MNRAS.319.1159W,2002MNRAS.330L..24S,Fan2002,2003ApJ...597..455K,2003ApJ...595..950Z}. The quiet periods of the prompt gamma-ray/X-ray emission have also been used to set upper limits on some GRB material \citep{Zou2010}. These approaches can be used to measure the ``averaged" Lorentz factor of the total GRB outflow material.

All of these methods have their own disadvantages, such as the ambiguity on variability timescale, dependence on other uncertain quantities (for example, the radiation efficiency), and the assumption on the microphysical parameters as well as the environment; it is worthwhile to compare the Lorentz factors derived in different ways. For such a purpose,  at least two observational features (i.e. high energy cutoff, thermal component, the well-behaved rising of the forward shock afterglow or the distinct reverse shock optical flash) are needed for the same event. Such a request is unsatisfied in most cases. In this work, we study one specific case$-$GRB 160625B, a burst is so bright that the spectrum can be well measured in very-short time intervals and the Lorentz factors of the fireball shells can be derived in a few approaches. In \S2, we perform the spectrum analysis of GRB 160625B. In \S3  we calculate the Lorentz factors from high energy cutoffs and the possible thermal component found in GRB 160625B, and compare them in the $\Gamma-L_{\gamma}$ relation with other bursts (where $L_{\gamma}$ represents the luminosity of the prompt emission). In \S4 we summarize our results with some discussions.

\section{Data Analysis} \label{sec:data}

\subsection{Observations} \label{subsec:obs}

GRB 160625B first triggered Fermi GBM at 22:40:16.28 UT on 25 June 2016 \citep{2016GCN...19581..1B}. About 188s later Fermi/LAT was triggered by a bright pulse from the same GRB and the onboard location is RA, Dec = 308.3, 6.9 (J2000) \citep{2016GCN...19580..1D}. This pulse accompanied very bright pulses seen by GBM. The Fermi GBM was triggered at 22:51:16.03 UT for the second time for this burst \citep{2016GCN...19581..1B}. Other gamma-ray telescopes, including Konus-Wind \citep{2016GCN...19604..1S} and CALET \citep{2016GCN...19597..1Y} also reported the detection of GRB 160628B. Swift/XRT has performed follow-up observations of this burst \citep{2016GCN...19585..1M}, and to date an afterglow of $\sim 10^6$ s  has been detected. There are also fruitful optical observations on the afterglow \citep{g1,g2,g3,g4,g5,g6,g7,g8,g9,g10,g11}, Xu et al. reported a redshift of 1.406 measured by the VLT/X-shooter \citep{g4}, which was then confirmed by TNG \citep{g5}. The isotropic-equivalent energy corresponding to this redshift is $\sim 5\times10^{54}$ erg in Konus-Wind's energy band \citep{2016GCN...19604..1S}. The afterglow of GRB 160625B is also detected on near infrared(NIR) and 15 GHz radio band \citep{2016GCN...19602..1W,2016GCN...19610..1M}.

In this work we mainly focus on analyzing the gamma-ray data from Fermi satellite to investigate the properties of prompt emission, yet we will also have discussion about results from other observations.

\subsection{Data Selection} \label{subsec:select}

We extract the GBM data, the standard LAT data as well as the LAT Low Energy (LLE) data of GRB 160625B from the Fermi Science Support Center (FSSC). For GBM data, we choose three NaI detectors that have the smallest angles from individual detectors boresight to the GRB when the burst was triggered, and the choice of BGO detector is based on the position corresponding to the selected NaI detectors. We use the Time-tagged Events (TTE) data files which contain individual photons with time and energy tags, and they cover a time range from $\sim 140$ s before $T_{\rm0}$ (the GBM first trigger time) to $\sim 480$ s after $T_{\rm0}$. For LLE data, we use the FITS file generated by LAT Low-Energy Events Catalog Server. It contains the events passing the LLECUT and has already been binned in energy and time, with 1 second resolution and covers a time range of $(-1000\ s, 1000\ s)$ with respect to $T_{\rm0}$. In the following joint spectral analysis we combine LLE data (30 MeV to 1 GeV) and the GBM data (10 keV to 800 keV for NaI detectors and 200 keV to 40 MeV for BGO detector) in the fitting.

\subsection{Spectral Fitting} \label{subsec:fit}

Herein we perform the joint spectral analysis using RMFIT version 4.3.2.  Our aim is to extract the time-resolved spectra and to search for the potential cutoff and black body components in prompt emission; the reasons and criterions for the division of time intervals are described as follows:

First, the time intervals of each spectrum should have strong gamma-ray signal detected over the background for all the selected detectors at the same time. Since there are much less photons at high energy end, the time division mainly depends on the LLE data. We set the minimum count rate for LLE data to be 40 counts/s, and this lead to a time span of $186-203$ s for GRB 160625B with respect to $T_{\rm0}$. Second, we divide the time span into several intervals: initially, we divide the time span into 1 s bins, then we combine the bins which have less than 200 LLE counts to the next one. Following these steps, we finally get 8 time intervals(see the first column of Table 1).

Initially, we fit each spectrum with Band function \citep{1993ApJ...413..281B} as the baseline. Then, we add high energy cutoff and black body components into the model to see how they improve the fit. We consider the cutoff on the high energy end of band function, therefore, we have the following models for comparison:

\begin{enumerate}
  \item Band function \\
  \[\mathop N\nolimits_{{\rm{band}}}  = \left\{ {\begin{array}{*{20}{c}}
  {A{{\left( {E/100} \right)}^\alpha }\exp \left( { - E\left( {2 + \alpha } \right)/{E_{{\rm{peak}}}}} \right)}&{{\rm{if\ }}E < {E_{\rm{b}}}}\\
  {A{{\left\{ {\left( {\alpha  - \beta } \right){E_{{\rm{peak}}}}/\left[ {100\left( {2 + \alpha } \right)} \right]} \right\}}^{\left( {\alpha  - \beta } \right)}}\exp \left( {\beta  - \alpha } \right){{\left( {E/100} \right)}^\beta }}&{{\rm{if\ }}E \ge {E_{\rm{b}}}}
\end{array}} \right.\ ,\]
where \[{E_{\rm{b}}} = \left( {\alpha  - \beta } \right){E_{{\rm{peak}}}}/\left( {2 + \alpha } \right)\ ,\]
  \item BandC model, i.e. the Band function with a high energy cutoff \\
  \[{N_{{\rm{BandC}}}} = {N_{{\rm{Band}}}}\exp \left( { - E/{E_{\rm{c}}}} \right)\ ,\]
  \item Band+BB, Band with a black body component \\
  \[{N_{{\rm{Band + BB}}}} = {N_{{\rm{Band}}}} + {A_{\rm{3}}}\frac{{{E^2}}}{{\exp \left( {E/kT} \right) - 1}}\ ,\]
  \item BandC + BB \\
  \[{N_{{\rm{BandC + BB}}}} = {N_{{\rm{BandC}}}} + {A_{\rm{3}}}\frac{{{E^2}}}{{\exp \left( {E/kT} \right) - 1}}\ ,\]
\end{enumerate}

The uncertainties caused by inter-calibration between the GBM and the LAT are taken into account, by adding an Eff. Area Corr. term in RMFIT \citep{2013ApJS..209...11A}. The correction factors are allowed to vary from 0.9 to 1.2 for NaI and BGO detectors, while fixed to 1 for LAT LLE.

The Castor Statistic ($CSTAT$) is chosen as the fitting statistic, since it is suitable for Poisson data, which is the case for energy bins at the high end.

We summarize the $CSTAT$ of the four models for each of our spectra in Table 1, and will discuss about the results in the next section.

\subsection{Fitting Result} \label{subsec:res}

As described in the previous section, we fit the time-resolved spectra of GRB 160625B with four different models. These models have different numbers of free parameters. In general, introducing more parameters will improve the fit, but one should also be aware that a complex model may over fit the data. To judge the most appropriate models that felicitously describe the data, we introduce the Bayesian information criterion ($BIC$). The $BIC$ was developed by Gideon E. Schwarz and is used to make selection among a finite set of models . The models being compared need not to be nested, and the model with the lowest $BIC$ is preferred. The $BIC$ is defined as \citep{1978AnSta...6..461S}:

\begin{displaymath}
  BIC =  - 2\ln {\cal L} + k\ln \left( N \right)\
\end{displaymath}
where $\cal{L}$ is the likelihood of the best-fit model, $k$ is the number of free parameters and $N$ is the number of data points respectively. The $CSTAT$ in our fit can be converted to likelihood by $CSTAT = -2ln\cal{L}$. We compute the $BIC$ for the four models and list them in Table 1. When comparing a model against another model with higher $BIC$, $\Delta BIC$ of $2-6$ represents a positive evidence, $\Delta BIC$ of $6-10$ represents a strong evidence and $\Delta BIC > 10$ represents a very strong evidence of improvement \citep{1998wst}. We note that the Bayesian Information Criterion compares models from pure statistical perspective, one should also consider the physical interpretation of models and the reasonable range of their parameter values.

By applying the criterion described above, we find that the fits are improved significantly ($\Delta BIC >10$) after adding extra components (i.e. the cutoff or black body components) into the model comparing to fitting the spectra with band function alone for all of the 8 time intervals. The $\rm{BandC+BB}$ model has the lowest $BIC$ in 6 intervals, while the BandC model and $\rm{Band+BB}$ model have the lowest $BIC$ in $186-188 \rm s$ and $191-195 \rm s$ respectively. During $201-203$ s, although the $\rm BandC+BB$ model has the lowest $BIC$, it is comparable with the $\rm Band+BB$ model ($\Delta BIC < 2$), and the cutoff energy is poorly constrained ($295\pm175$ MeV). Thus, we prefer the $\rm{Band+BB}$ model rather than the $\rm BandC+BB$ model to represent the spectral shape for this interval. Extra power law components are found in some GRBs, for completeness, we have also included the models with this component in the comparison; however, it didn't improve the fit significantly.

As mentioned above, the $BIC$ compares models from pure statistical perspective. Although adding a thermal component into the model significantly improves the fit in 7 of 8 time intervals, we do not claim a clear detection of thermal component in GRB 160625B for the following reasons: firstly, the thermal components are just sub-dominant in all of the 7 intervals (i.e., they just account for $\sim 14-28\%$ total luminosities); secondly, there may be strong spectral evolution within a time scale of 1 second (which is the smallest scale of our time bins that is limited by the LLE data). The superposition of Band functions with different $E_{\rm peak}$ may also lead to a variant on the shape of the time-average spectrum. To further examine the presence of thermal components, we divided two brightest time intervals ($188-189 \rm s$ and $189-190 \rm s$) into ten 0.2 s bins, and fit the data of n9 and b1 detectors (We ignore the LLE data, since the LLE data of this burst are insufficient for such a short time bin, and the possible thermal components are not within the energy range of LLE data) with Band and $\rm{Band+BB}$ models. The result shows that in 3 bins there are very strong evidence, in two bins we have strong evidence and in another two bins we have positive evidence of improvement on the fit after adding the black body component according to the $BIC$. The other 3 bins are the least bright ones, and their BGO data above 2000 keV are mostly upper limits, so it is hard to constrain an extra component located on the high energy end of Band function.

To summarize the fitting result of GRB 160625B, we found cutoffs of tens of MeV in 6/8 of the intervals, and the evidence of thermal radiation  in 7/8 of the intervals; the evidence of thermal radiation component still exists even in 0.2 s resolution for the two brightest intervals. We present the models we prefer for the 8 intervals in Table.1, and our following calculations are based on the parameters of these preferred models listed in Table 2.

\section{Model-dependent estimates of the Lorentz factors of the outflow material} \label{sec:disc}

In this section we compute the Lorentz factors using the high energy cutoffs and thermal radiation components that obtained in section 2, then test the correlation between the Lorentz factor and the rest frame isotropic gamma-ray luminosity ($\Gamma-L_{\gamma}$) using our results and compare them with other works.

\subsection{Evaluating the Lorentz factor} \label{subsec:dtlf}

\subsubsection{The Cutoff model} \label{subsec:lcut}

For GRB 160625B, since we have found high energy cutoffs in time-resolved spectra in 6 intervals, we can calculate the Lorentz factors for these intervals and explore how they evolve. Assuming the cutoffs are caused by the $\gamma\gamma$ absorption, then the Lorentz factor $\Gamma$ can be derived by \citet{2001ApJ...555..540L}:

\begin{equation}\label{lcut}
  \Gamma  = {\hat \tau ^{1/\left( { - 2\beta  + 2} \right)}}{\left( {{E_{\rm{c}}}/{m_{\rm{e}}}{c^2}} \right)^{\left( { - \beta  - 1} \right)/\left( { - 2\beta  + 2} \right)}}{\left( {1 + z} \right)^{\left( { - \beta  - 1} \right)/\left( { - \beta  + 1} \right)}},
\end{equation}
where $\beta$ is the high energy index of Band function, $z$ is the redshift, and $E_{\rm{c}}$ is the cutoff energy. Numerically, $\hat \tau$ can be calculated by \citet{2001ApJ...555..540L}
\begin{displaymath}
  \hat \tau  = \left( {2.1 \times {{10}^{11}}} \right)\left[ {\frac{{{{\left( {d_{\rm L}/7{\rm{Gpc}}} \right)}^2}{{\left( {0.511} \right)}^{\beta  + 1}}{f_{\rm{1}}}}}{{\left( {\delta T/0.1{\rm{s}}} \right)\left( { - \beta  - 1} \right)}}} \right],
\end{displaymath}
where $f_{\rm1}$ is the observed number of photons per second per square centimeter per MeV at the energy of 1 MeV, and $\delta T$ is the variability timescale.

We find that the cutoff energies are relatively low, $\sim$ tens of MeV. When using eq.(1) to derive the Lorentz factor, it is assumed that the photons with energy $E_{\rm c}$ can annihilate a second photon whose energy is much less than $E_{\rm c}$, i.e., ${E_{\rm{c}}} \gg {\Gamma ^2}m_{\rm{e}}^2{c^4}/\left[ {{E_{\rm{c}}}{{\left( {1 + z} \right)}^2}} \right]$. If this is not satisfied, eq.(1) is no longer valid since the spectrum of target photons cannot be described by a power law parameterized by $\beta$ (the high energy index of band function). In this case, we can only assume that the photons with energies around $E_{\rm c}$ annihilate with target photons with energies comparable to themselves, and then the Lorentz factor is estimated by
\begin{equation}\label{lcut2}
  \Gamma  \approx \frac{{{E_{\rm{c}}}}}{{{m_{\rm{e}}}{c^2}}}\left( {1 + z} \right).
\end{equation}
The Lorentz factors derived from the opacity hypothesis are shown in Table 3. However, another limitation on Lorentz factor should be considered when $E_{\rm c}$ is low. The electron-positron pairs that produced by photon annihilation can in turn Compton scatter other photons, which sets a lower limit on Lorentz factor \citep{2001ApJ...555..540L}
\begin{displaymath}
  \Gamma  > {\hat \tau ^{1/\left( { - \beta  + 3} \right)}}{\left( {1 + z} \right)^{\left( { - \beta  - 1} \right)/\left( { - \beta  + 3} \right)}}{\left( {180/11} \right)^{1/\left( {6 - 2\beta } \right)}},
\end{displaymath}
if this limit is unsatisfied, the burst would be optically thick to all photons \citep{2001ApJ...555..540L}. We calculate this limit for all of our spectra with cutoff and list them in Table 3. Surprisingly, the lower limits are much (about an order of magnitude) higher than the Lorentz factors derived from opacity hypothesis for GRB 160625B. The inconsistency of the results from the two methods implies that the spectral cutoffs of GRB 160625B are unlikely caused by pair production of high energy photons. There is an additional argument disfavoring the absorption hypothesis. As found in the numerical simulations \citep[e.g.,][]{2004ApJ...613..448P}, in order to have an exponential cutoff due to
absorption, one would need an absorbing screen through which the radiation propagates.
While in reality the absorption and emission processes are coexisting. With a
proper radiation transfer the observed absorption feature is a break in the
power-law slope, not an exponential cutoff \citep{2004ApJ...613..448P}.

\subsubsection{The Thermal radiation model} \label{subsec:lbb}

The measurements of the temperature and flux of the thermal components also allow the determination of the fireball shells' Lorentz factor. We evaluate the Lorentz factor by \citet{2007ApJ...664L...1P}
\begin{equation}\label{lbb}
  \Gamma  = {\left[ {\left( {1.06} \right){{\left( {1 + z} \right)}^2}{d_{\rm{L}}}\frac{{Y{F_{\gamma {\rm{,ob}}}}{\sigma _{\rm{T}}}}}{{2{m_{\rm{p}}}{c^3}{\cal R}}}} \right]^{1/4}},
\end{equation}
where $F_{\gamma {\rm{,ob}}}$ is the observed total energy flux, $Y$ is the ratio between the total fireball energy and the energy emitted in gamma-rays, and ${\cal R}$ is defined as
${\cal R} \equiv {\left( {\frac{{{F_{{\rm{bb,ob}}}}}}{{\sigma T_{{\rm{ob}}}^4}}} \right)^{1/2}}$,
where $F_{\rm bb,ob}$ and $T_{\rm ob}$ are the observed blackbody component flux and temperature respectively. Meanwhile, three relevant radius -- the initial fireball radius $r_{\rm 0}$, the saturation radius $r_{\rm s}$, and the photospheric radius $r_{\rm ph}$ can be obtained by \citet{2007ApJ...664L...1P}:
\begin{equation}\label{r0}
  {r_{\rm{0}}} = \frac{{{4^{3/2}}}}{{{{\left( {1.48} \right)}^6}{{\left( {1.06} \right)}^4}}}\frac{{{d_{\rm{L}}}}}{{{{\left( {1 + z} \right)}^2}}}{\left( {\frac{{{F_{{\rm{bb,ob}}}}}}{{Y{F_{{\rm{\gamma,ob}}}}}}} \right)^{3/2}}R \ ,\
\end{equation}
\begin{equation}\label{rs}
  {r_{\rm{s}}} = \Gamma {r_{\rm{0}}}\ ,\
\end{equation}
\begin{equation}\label{rs}
  {r_{{\rm{ph}}}} = {E_{{\rm{tot}}}}{\sigma _{\rm{T}}}/8\pi {\Gamma ^3}{m_{\rm{p}}}{c^3}\ .\
\end{equation}
where the total energy of the fireball $E_{\rm total}$ can be estimated by ${E_{{\rm{total}}}} = \pi d_{\rm{L}}^2Y{F_{{\rm{\gamma,ob}}}}$. With eqs.(3-6), we calculate $\Gamma$, $r_{\rm 0}$, $r_{\rm s}$, and $r_{\rm ph}$ for the 7 intervals of GRB 160625B that are likely to host thermal components, and the results are also summarized in in Table 3. For these calculations, it is assumed that the blackbody components in different intervals are dominated by thermal emissions from independent shells, and the high latitude emission from the previous interval is not considered. Note that in the above approach we adopt an analytic approximation that the outflow accelerates linearly at first, and then moves in a constant speed after reaching the saturation radius. The actual transition could be much smoother \citep[see][and the references therein]{1999PhR...314..575P}, likely affecting the estimates of Lorentz factor and the nozzle radius. In the current scenario usually we have $r_{\rm ph}\sim {\rm a~few}\times r_{\rm s}$ for a reasonable $Y\sim 4$ (see Table 3) and hence the analytic approximation seems reasonable.

We find that the $\Gamma$ derived from blackbody components distribute from $900$ to $2000$, which are much higher than that derived from the spectral cutoffs of tens of MeV, and is satisfied with the limitation set by Compton scattering effect. This again suggests the cutoffs in GRB 160625B are not caused by the pair production effect. An upper limit of central engine¡¯s mass can be set by assuming the initial fireball radius is (of course) outside the Schwarzschild radius of the central black hole, then the upper limit is derived by $m < \left( {{r_{\rm{0}}}{c^2}/2G} \right)Y^{-3/2}$. We plot the upper limits derived from different intervals of GRB 160625B with different $Y$ in Figure 2. If $Y$ is larger than 4 in the first two time intervals, the black hole's mass will be lower than $2M_\odot$, which is lower than the maximal gravitational mass of neutron stars measured so far.

\subsubsection{Correlations} \label{subsec:corr}

Correlations involve $\Gamma$ are widely discussed in the literature, since they give important clues to reveal the physics of GRB. \citet{2010ApJ...725.2209L} found a tight correlation between $\Gamma$ and isotropy gamma-ray energy $\Gamma  \propto E_{\gamma}^{0.25}$. \citet{2012ApJ...751...49L} extended the sample and found another tight correlation of $\Gamma  \propto L_{\gamma}^{0.3}$. Later, \citet{2012ApJ...755L...6F} showed that the time-resolved thermal emissions of GRB 090902B also follow the $\Gamma-L_{\gamma}$ correlation. We test this relation with our results, and also include the samples from \citet{2015ApJ...806..194T} in Figure 3. The grey points and grey solid line are samples from \citet{2012ApJ...751...49L} and the empirical correlation they derived respectively. We find that the Lorentz factors derived from the thermal components show a tight positive correlation with $L_{\gamma}$ (with Pearsons correlation coefficient of 0.91 and is irrelevant to the value of $Y$ in eq.(3)). The red triangles in Figure 3 are calculated with eq.(3) by setting $Y =1$ (corresponding to a very high radiation efficiency case), we find that the sequence in GRB 160625B is very similar to the one in GRB 090902B (blue triangles). Fitting these two sequences respectively, we obtain the slop of $0.40\pm0.08$ for GRB 160625B (red dash line) and $0.39\pm0.03$ for GRB 090902B (blue dash line), which are consist with each other within the errors. On the other hand, the data of GRB 160625B in Figure 3 also follow the sequence for different bursts obtained by \citet{2012ApJ...751...49L}. We note that although the slop they derived is $0.29\pm0.002$, the relatively large dispersion (with a Pearsons correlation coefficient of 0.79 \citep{2012ApJ...751...49L}) would lead to the change of slop for different group of samples.

The green dots in Figure 3 are derived from $\gamma\gamma$ opacity hypothesis which do not satisfy the lower limits (blue arrows) set by Compton scattering effect, and it is clear that they do not show a $\Gamma-L_\gamma$ correlation.

\subsubsection{Information from the afterglow} \label{subsec:lbb}

As mentioned in section 2, the follow up observations from radio to X-ray band can also be utilized to infer information about the outflow.

Swift/XRT began to observe the afterglow 10000 s after the triggered \citep{2016GCN...19585..1M}. Although the onset of the afterglow was not seen due to the relatively late start time of observation, one can still obtain a lower limit for Lorentz factor by requiring that the outflow is fast enough to produce the onset before the observation time. Assuming the afterglow of GRB 160625B is in the thin shell case and the environment is homogenous, the lower limit can be derived by \citep{1999A&AS..138..537S}
\begin{equation}\label{limitaf}
   \Gamma  > 193{\left( {n\eta } \right)^{ - 1/8}} \times {\left( {\frac{{{E_{\gamma {\rm{,52}}}}}}{{t_{{\rm{p,z,2}}}^3}}} \right)^{1/8}}.
\end{equation}
We collect the total fluence of GRB160625B from FSSC, and K-corrected \citep{2001AJ....121.2879B} it into the rest frame isotropic-equivalent gamma-ray energy $E_{\gamma}$ in $1-10000$ keV band. The $E_{\gamma}$ are found to be $9.20\pm0.02\times10^{54}$ erg, which is very high among GRBs. We take the radiation efficiency $\eta =0.5$ and the circumburst density $n = 0.1 \ cm^{-3}$. Let $t_{\rm{p,z,2}}$ equals to the start time of XRT observatiion in the rest frame, the lower limit derived from eq.(7) is 164.5. We note that the limitation here is for the bulk Lorentz factor of the merged shells that crashed into the surrounding medium, while the Lorentz factors or limitations derived from the previous sections are for the independent shells before they merged.

Another important phenomenon observed by Swift/XRT is the jet break of X-ray afterglow at late time. We use the light curve analysis result from UK Swift Science Data Centre, in which the jet break time is determined to be $1.8\pm0.5\times10^6~{\rm s}$ \citep{2007A&A...469..379E,2009MNRAS.397.1177E}. With the isotropic Energy $E_{\gamma}$ and the jet break time $t_{\rm j}$, the half-opening angle of the jet can be estimated by \citep{1999A&AS..138..537S,2001ApJ...562L..55F}:

\begin{equation}\label{limitaf}
  {\theta _{\rm{j}}} \approx 0.057{\left( {\frac{{{t_{\rm{j}}}}}{{1{\rm{ day}}}}} \right)^{3/8}}{\left( {\frac{{1 + z}}{2}} \right)^{ - 3/8}}{\left( {\frac{{{E_\gamma }}}{{{{10}^{53}}{\rm{ erg}}}}} \right)^{ - 1/8}}{\left( {\frac{\eta }{{0.2}}} \right)^{1/8}}{\left( {\frac{{{n}}}{{0.1{\rm{ c}}{{\rm{m}}^{ - 3}}}}} \right)^{1/8}}
\end{equation}
We still assume $\eta$ and $n$ to be 0.5 and 0.1 respectively, then we obtain $\theta_{\rm j} = 0.106 \ rad$. Having $\theta_{\rm j}$, the beaming-corrected energy can be calculated by ${E_{\gamma {\rm{,j}}}} = {E_\gamma }\left[ {1 - \cos \left( {{\theta _{\rm{j}}}} \right)} \right]$. We find that the beaming-corrected energy for GRB 160625B is extremely high, up to $\sim 5.15\times10^{52}$ erg. Considering the error of $t_{\rm j}$ and the uncertainties of $\rm \eta$ and $n$ (the error of the fluence is less than 1 percent and need not to be considered), we estimate the lower and upper limits for $\theta_{\rm j}$ and $E_{\gamma {\rm{,j}}}$ by setting the parameter set ($t_{\rm j},\eta,n$) in eq.(8) to be ($1.3\times10^6,~0.1,~0.001$) and ($2.3\times10^6,~0.9,~10$) respectively. The $\theta_{\rm j}$ and $E_{\gamma {\rm{,j}}}$ with lower and upper limits (treated as errors) computed in this way is then to be $E_{\gamma {\rm{,j}}} = 5.15^{+17.4}_{-4.29}\times10^{52}$ erg and $\theta_{\rm j} = 0.106^{+0.116}_{-0.063}$ rad. To compare GRB 160625B with other bursts, we collect the samples from Table 1 and Table 2 of \citet{2016ApJ...818...18G} and calculate their $E_{\gamma, {\rm j}}$. For simplicity, we calculate the luminosity distance using the redshift, assuming a flat universe with $\Omega_{\rm M} = 0.286$, $\Omega_\Lambda = 0.714$ and $H_{\rm 0} = 69.6$ \citep{2006PASP..118.1711W}, and take $\eta = 0.5$, $n = 0.1 ~{\rm cm^{-3}}$ for all of the bursts. We plot the $E_{\gamma {\rm{,j}}}$ distribution of these bursts (blue bars) in Figure 4, and GRB 160625B (solid line) as well as its lower and upper limits (red dash lines). We can find from Figure 4 that the $E_{\gamma {\rm{,j}}}$ of GRB 160625B is higher than any of the previous bursts under the typical parameters.

Combining the Swift/XRT data with the observation from optical and radio band collected from GCN, we also make an attempt to fit the multi-wavelength afterglow with the forward shock model, in which the Lorentz factor ($\Gamma$), the isotropic equivalent kinetic energy ($E_{\rm k}$), the circumburst density ($n$), microphysical parameters ($\varepsilon_{\rm e}$, $\varepsilon_{\rm b}$), spectral index of the electron energy distribution ($p$) as well as the half-opening angle ($\theta_{\rm j}$) are taken as free parameters. We consider a homogeneous environment, and the revised fireball dynamics proposed by \citet{1999MNRAS.309..513H} are used. With the code initially developed in \cite{Fan2006} we find that the parameter set of $\Gamma = 200$ (which should be taken as an lower limit since we just fitted the data at $t\geq 10^{4}$ s), $E_{\rm k} = 6\times10^{53}$ erg, $n = 0.07~{\rm cm}^{-3}$, $\varepsilon_{\rm e} = 0.3$, $\varepsilon_{\rm b} = 0.001$, $p = -2.1$ and $\theta_{\rm j} = 0.135$ can reasonably reproduce the late time (i.e., $t>10^{4}$ s) afterglow data, as shown in Figure 5. The half-opening angle and the circum-burst density are consistent with what we assumed to derive the beaming-corrected energy above (with this half-opening angle, $E_{\gamma {\rm{,j}}}\approx 8\times 10^{52}~{\rm erg}$), and $E_{\rm k} = 6\times10^{53}$ erg corresponds to an extremely high radiation efficiency of $\sim 94\%$. We note that due to the lack of a well-behaved rise of the afterglow, not all of the parameters can be well determined, especially the initial Lorentz factor. If what we obtained from the forward shock modeling (i.e., $\Gamma = 200$) is close to the real situation, the much lower bulk Lorentz factor comparing to the Lorentz factors of the unmerged shells can be explained by a great amount of kinetic energy of the fast shells has been transferred into the radiation.

At last, the Pi of the Sky Telescope has detected a very bright optical flare accompanying GRB 160625B \citep{g10}. More efforts are needed to identify the origin of this flash (one possibility is that such a flash was triggered by the main GRB outflow ejected at $t\sim 186$ s catching up with the decelerated outflow material ejected at $t\sim 0$ s). If it was originated from the reverse shock of the outflow, the Lorentz factor can be measured in another way for GRB 160625B \citep{1999A&AS..138..537S,2000MNRAS.319.1159W,2002MNRAS.330L..24S,Fan2002,2003ApJ...597..455K,2003ApJ...595..950Z}. The detailed modeling of such a component is beyond the scope of this work.

\section{Conclusion} \label{sec:con}
In the literature several methods have been proposed to estimate the Lorentz factors of the GRB outflow material. Some methods are only applicable to the whole burst (for example, the methods based on the reverse shock optical flash modeling or the forward shock emission rise modeling), while some methods are valid for the time-resolved outflow material. The robustness of these estimates should be cross checked. In reality such a goal is however hard to achieve due to the limited data. In this work we show that GRB 160625B, an extremely-bright long GRB with well-measured spectrum, provides us the valuable chance to do that. We perform spectral analysis on GRB 160625B, and find cutoffs and the evidence of thermal components in its time-resolved spectra. The cutoffs in GRB 160625B are unlikely caused by pair productions, since the Lorentz factors derived from the cutoffs are well below the lower limits set by Compton scattering effect (see Figure 3). Instead these cutoffs may trace the spectra of accelerated electrons. The Lorentz factors derived from the thermal components within GRB 160625B follow the $\Gamma-L_{\gamma}$ correlation with a slop of 0.40, which is nicely consistent with that holds for time-resolved distinct thermal components of GRB 090902B \citep{2012ApJ...755L...6F}. The consistence between the correlations found in GRB 160625B and GRB 090902B strengthens the presence of the thermal radiation components in GRB 160625B. In view of these facts, {\it there is the caution on estimating the Lorentz factors of the GRB outflow solely with the  cutoff(s) in the energy spectra, in particular the cutoffs appeared at low-energies (i.e., $\sim 10{\rm s}$ MeV, as found in our cases).}

We calculate the upper limits on the mass of the central black hole of GRB 160625B for different $Y$, and find that $Y$ should not be larger than $\sim 4$ in the first two time-intervals displaying thermal signature, otherwise the mass of the black hole will be lower than $2M_\odot$, which has been ruled out by the latest neutron star mass measurement in which the lower limit on the maximal gravitational mass is $2.01\pm0.04~M_\odot$ \citep{Antoniadis2013}. Interestingly, as shown in Figure 2 there might be evidence for the increases of the mass of the central black hole. Indeed the main outburst starting at $t\sim 186$ s may be due to the formation of a black hole. The {\it extremely high geometry-corrected prompt gamma-ray energy $E_{\gamma, \rm j}\sim 5\times 10^{52}~{\rm erg}$ (or even $E_{\gamma, \rm j}\sim 8\times 10^{52}$ erg if we adopt the half-opening angle found in the numerical modeling of the late time afterglow, see Figure 5)} is also in support of the black hole central engine, while a magnetar with a spin period $\lesssim 1~{\rm ms}$ and a typical moment of inertia $I\sim 2\times 10^{45}~{\rm g~cm^{2}}$ seems hard to reproduce the data.

\acknowledgements
We thank the referee for helpful comments and Prof. E. W. Liang and Dr. B. B. Zhang for helpful communications/discussions. This work was supported in part by the National Basic Research Program of China (No. 2013CB837000 and No. 2014CB845800), the National Natural Science Foundation of China under grants No. 11525313 (that is the Funds for Distinguished Young Scholars), 11273063, 11433009 and 11303098, the Chinese Academy of Sciences via the External Cooperation Program of BIC (No. 114332KYSB20160007).

\clearpage

\begin{deluxetable}{CCCCCCc}
\tablenum{1}
\tablecaption{Comparison of the goodness of fit for different models\label{tab:lf}}
\tablewidth{0pt}
\tablehead{
\colhead{Time} & \colhead{Band} &
\colhead{BandC} &
\colhead{Band+BB} & \colhead{BandC+BB} &
\colhead{Preferred} \\
\colhead{(s)} & \colhead{} & \colhead{$CSTAT(BIC)$}
& \colhead{} & \colhead{} & \colhead{model}
}
\startdata
186-188	&	573.9 	 (	598.5 	) &	464.1 	 (	494.8 	) &	499.7 	 (	536.7 	) &	461.5 	 (	504.6 	) &	\rm BandC	\\
188-189	&	841.0 	 (	865.6 	) &	588.0 	 (	618.8 	) &	708.6 	 (	745.5 	) &	544.4 	 (	587.5 	) &	\rm BandC+BB	\\
189-190	&	904.2 	 (	928.7 	) &	759.5 	 (	790.2 	) &	696.4 	 (	733.3 	) &	644.9 	 (	687.8 	) &	\rm BandC+BB	\\
190-191	&	672.1 	 (	696.6 	) &	633.9 	 (	664.5 	) &	607.6 	 (	644.3 	) &	573.0 	 (	615.9 	) &	\rm BandC+BB	\\
191-195	&	826.2 	 (	850.9 	) &	826.3 	 (	857.1 	) &	698.2 	 (	735.2 	) &	697.7 	 (	740.9 	) &	\rm Band+BB	\\
195-200	&	828.2 	 (	852.9 	) &	820.9 	 (	851.7 	) &	751.1 	 (	788.1 	) &	716.5 	 (	759.7 	) &	\rm BandC+BB	\\
200-201	&	728.0 	 (	752.6 	) &	577.9 	 (	608.6 	) &	621.6 	 (	658.5 	) &	560.1 	 (	603.2 	) &	\rm BandC+BB	\\
201-203	&	606.3 	 (	630.9 	) &	604.5 	 (	635.2 	) &	578.5 	 (	615.4 	) &	570.7 	 (	613.8 	) &	\rm Band+BB	\\
\enddata

\end{deluxetable}

\begin{deluxetable}{cCccccc}
\tablenum{2}
\tablecaption{Best--fit parameters of the assumed models\label{tab:para}}
\tablewidth{0pt}
\tablehead{
\colhead{Time} & \colhead{Model} & \colhead{$\alpha$} & \colhead{$\beta$} & \colhead{$E_{\rm peak}$} & \colhead{Temperature} & \colhead{$E_{\rm c}$} \\
\colhead{(s)} & \colhead{} & \colhead{} & \colhead{} & \colhead{(keV)} & \colhead{(keV)} & \colhead{(MeV)}
}
\startdata
186--188 & \rm BandC    & -0.79$\pm$0.01 & -1.61$\pm$0.02 & 1091$\pm$69 &       --       & 15.26$\pm$2.50 \\
188--189 & \rm BandC+BB & -0.54$\pm$0.01 & -1.75$\pm$0.01 & 572$\pm$17  & 389.0$\pm$7.92 & 19.15$\pm$2.45 \\
189--190 & \rm BandC+BB & -0.60$\pm$0.01 & -2.31$\pm$0.01 & 613$\pm$10  & 657.3$\pm$10.8 & 58.77$\pm$18.4 \\
190--191 & \rm BandC+BB & -0.59$\pm$0.01 & -2.16$\pm$0.02 & 353$\pm$9   & 290.5$\pm$6.32 & 41.43$\pm$13.5 \\
191--195 & \rm Band+BB  & -0.60$\pm$0.01 & -2.58$\pm$0.02 & 320$\pm$10  & 243.1$\pm$10.4 &       --       \\
195--200 & \rm BandC+BB & -0.60$\pm$0.01 & -2.38$\pm$0.02 & 319$\pm$6   & 206.1$\pm$1.95 & 34.08$\pm$10.1 \\
200--201 & \rm BandC+BB & -0.60$\pm$0.01 & -1.91$\pm$0.02 & 459$\pm$17  & 250.4$\pm$5.30 & 32.53$\pm$5.74 \\
201--203 & \rm Band+BB  & -0.66$\pm$0.02 & -2.56$\pm$0.03 & 416$\pm$25  & 260.2$\pm$16.1 &       --       \\
\enddata

\end{deluxetable}

\begin{deluxetable}{ccccccccc}
\tablenum{3}
\tablecaption{Physical parameters derived from spectral analysis\label{tab:lf}}
\tablewidth{0pt}
\tablehead{
\colhead{Time} & \colhead{$L_{\rm \gamma,52}$} &
\colhead{$\Gamma _{\rm Cut}$} & \colhead{$\Gamma _{\rm BB}/Y^{1/4}$} &
\colhead{$\Gamma _{\rm limit}$} & \colhead{$r_{\rm 0}/Y^{-3/2}$} &
\colhead{$r_{\rm s}/Y^{-5/4}$} & \colhead{$r_{\rm ph}/Y^{1/4}$} & \colhead{$M_{\rm limit}/Y^{-3/2}$} \\
\colhead{(s)} & \colhead{($10^{52}\rm{erg}$)} & \colhead{} & \colhead{} & \colhead{}
& \colhead{($10^{7}\rm{cm}$)} & \colhead{($10^{10}\rm{cm}$)} & \colhead{($10^{11}\rm{cm}$)} & \colhead{($M_\odot$)}
}
\startdata
188--189	&	84.52	&	90.17$\pm$11.54	&	1656.35	&	924.78	&	0.52	&	0.86	&	2.63	&	17.65	\\
189--190	&	78.43	&	276.71$\pm$86.63	&	2022.59	&	485.49	&	0.41	&	0.83	&	1.34	&	13.94	\\
190--191	&	32.48	&	195.07$\pm$63.56	&	1215.12	&	462.60	&	1.18	&	1.43	&	2.56	&	39.80	\\
191--195	&	18.60	&	--	 &	1018.72	&	--	&	1.68	&	1.71	&	2.49	&	57.00	\\
195--200	&	21.87	&	160.46	$\pm$	47.55	&	933.71	&	337.52	&	3.78	&	3.53	&	3.80	&	127.87	\\
200--201	&	58.61	&	153.16	$\pm$	27.03	&	1274.04	&	695.07	&	0.99	&	1.26	&	4.01	&	33.48	\\
201--203	&	25.65	&	--	 &	1113.60	&	--	&	1.36	&	1.51	&	2.63	&	46.04	\\
\enddata
\tablecomments{$\Gamma _{\rm Cut}$,$\Gamma _{\rm BB}$,$\Gamma _{\rm limit}$ and $M_{\rm limit}$ represent the Lorentz factor derived from the cutoffs, the Lorentz factor derived from the thermal component, the lower limit of Lorentz factor derived from Compton scattering effect, and the lower limit for the mass of central black hole respectively.}
\end{deluxetable}

\clearpage

\begin{figure}[ht!]
\figurenum{1}
\centering
\includegraphics[angle=0,scale=0.3]{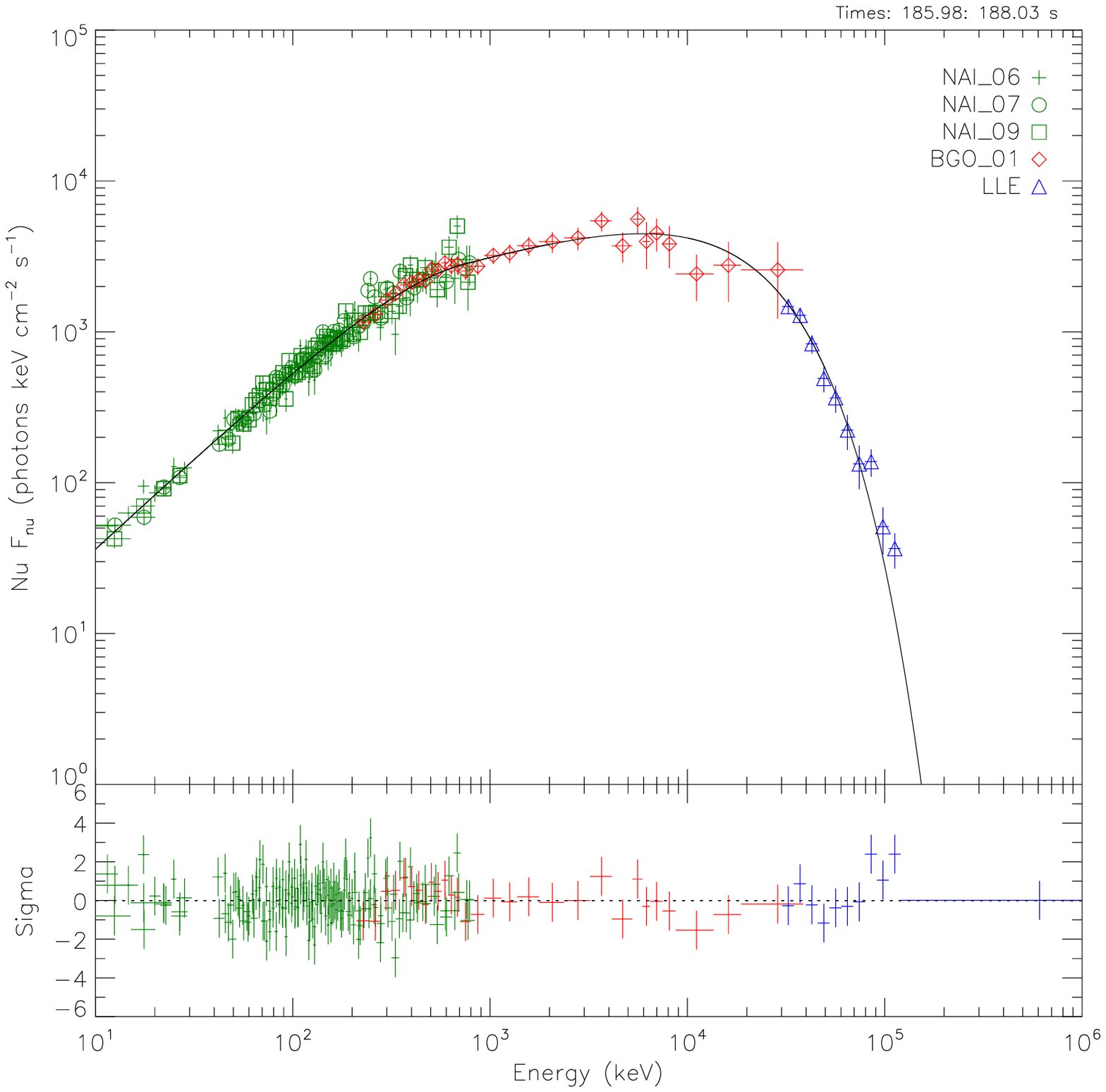}
\includegraphics[angle=0,scale=0.3]{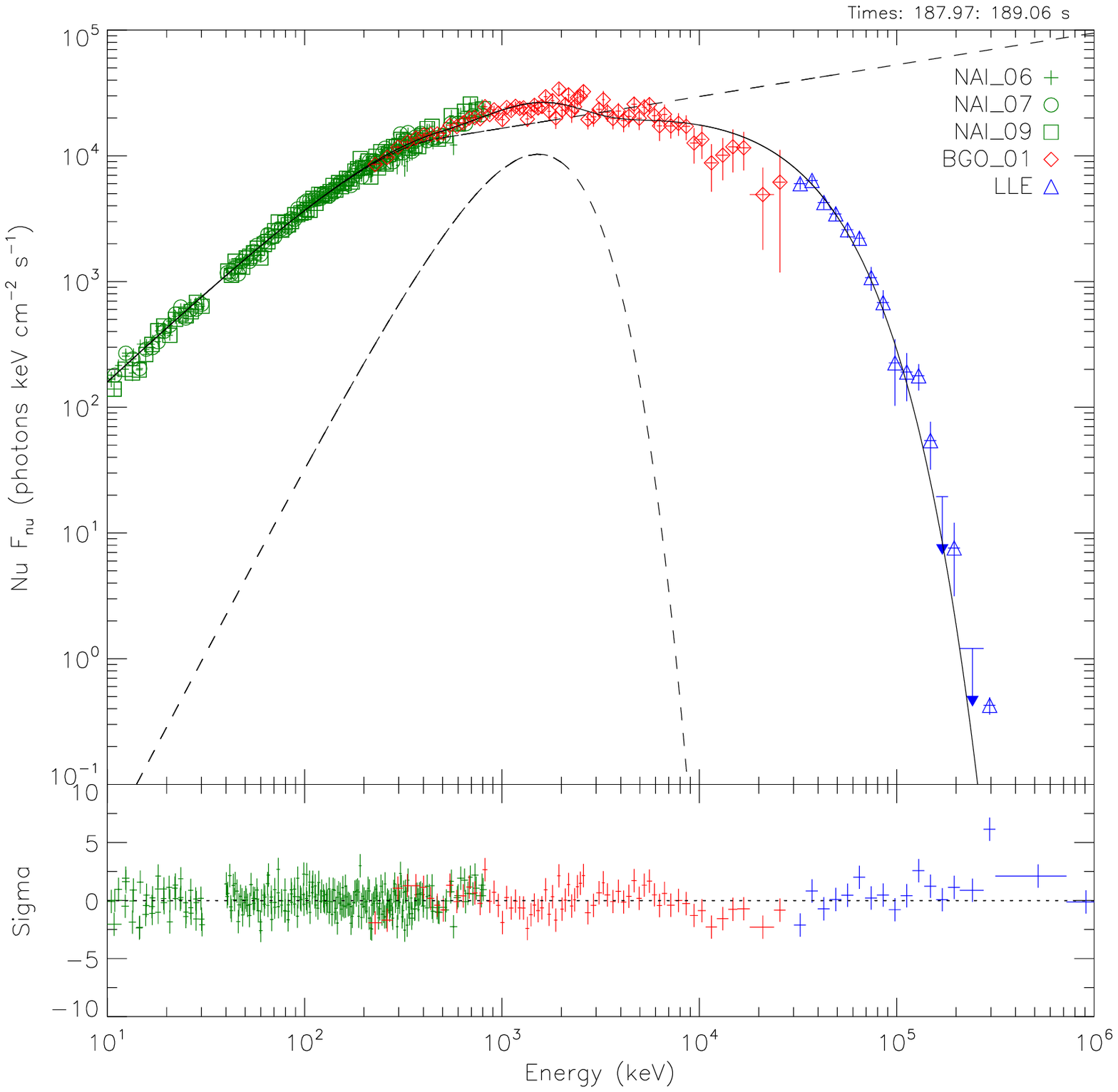}
\includegraphics[angle=0,scale=0.3]{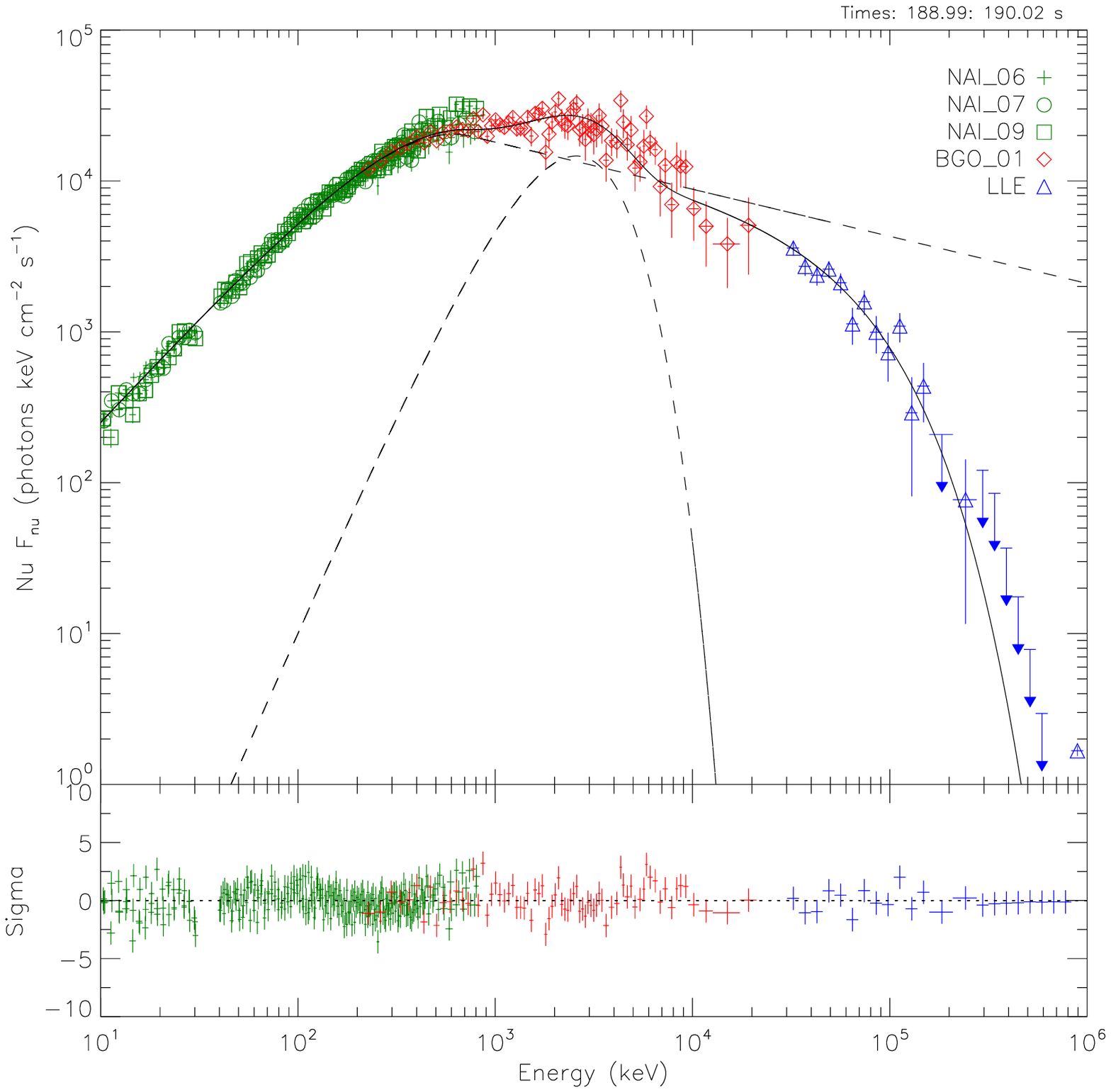}
\includegraphics[angle=0,scale=0.3]{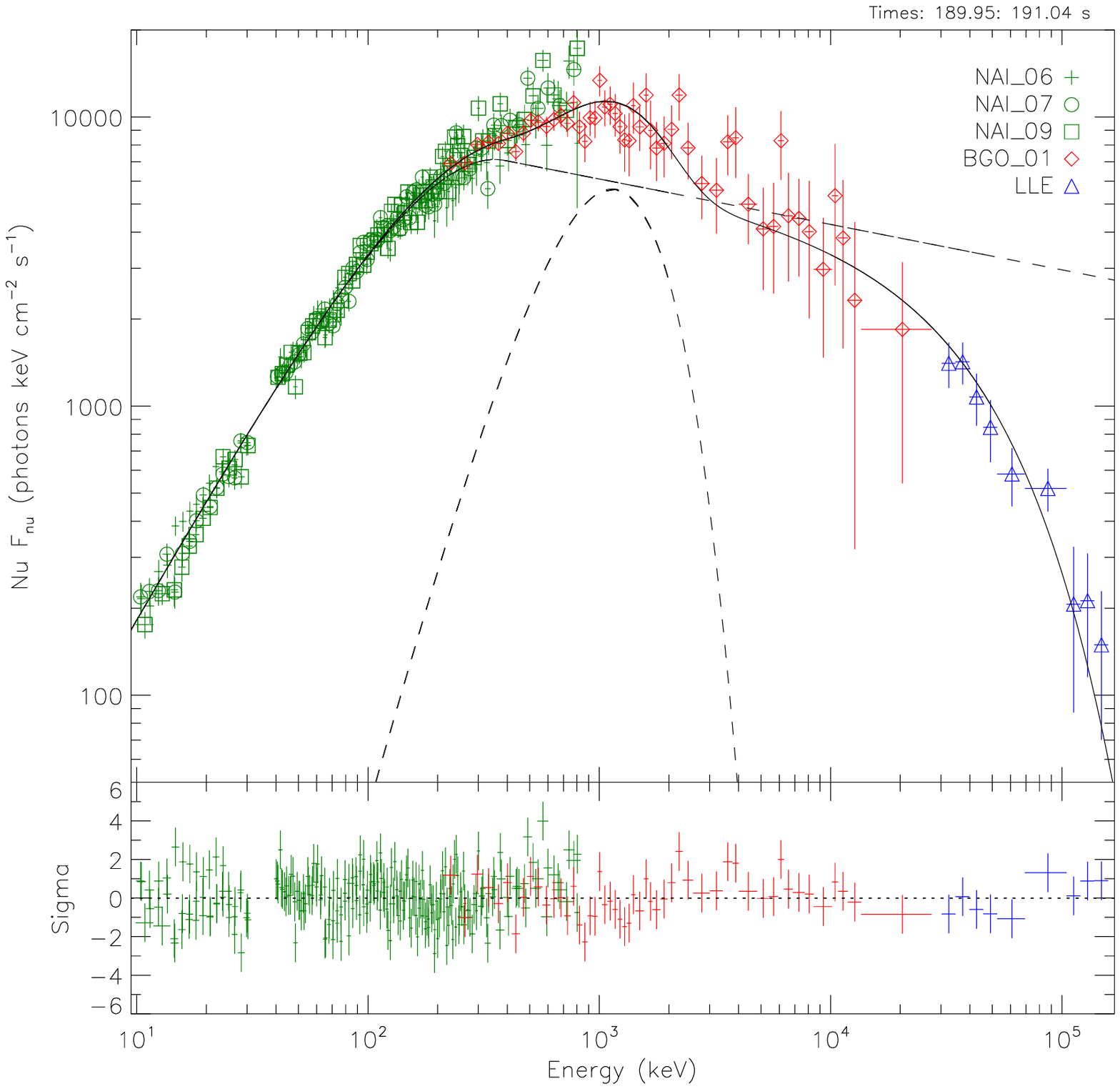}
\includegraphics[angle=0,scale=0.3]{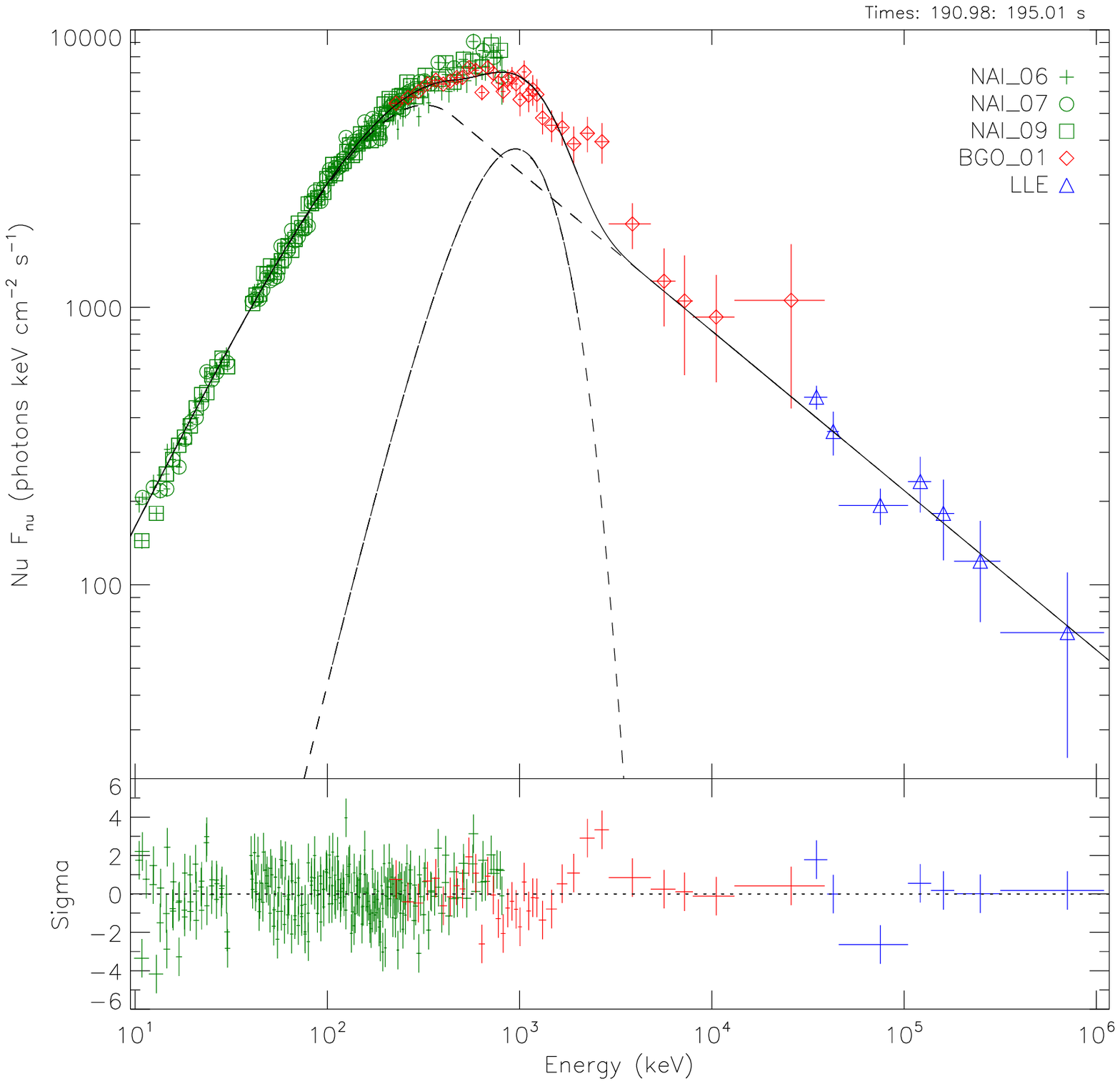}
\includegraphics[angle=0,scale=0.3]{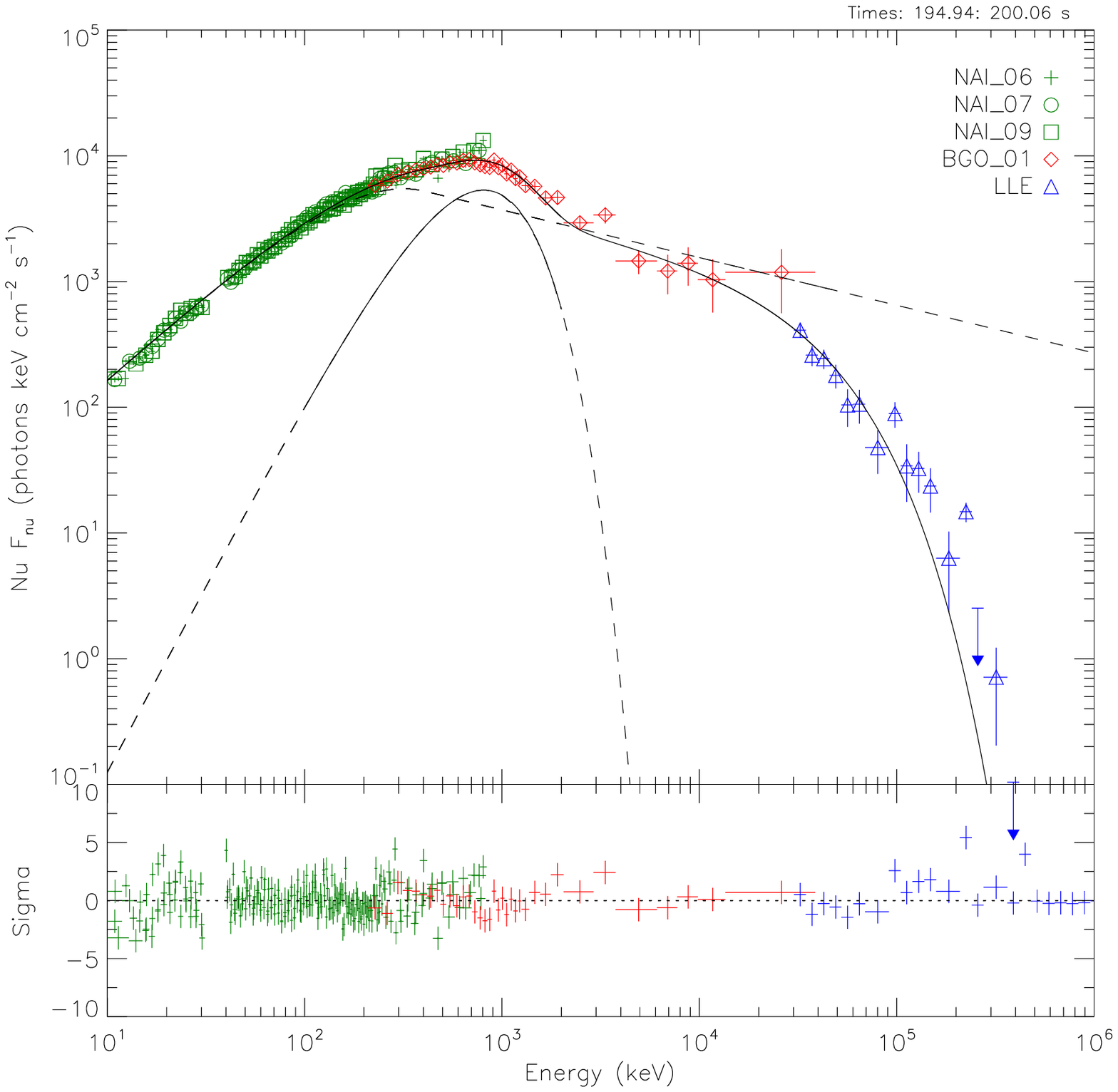}
\includegraphics[angle=0,scale=0.3]{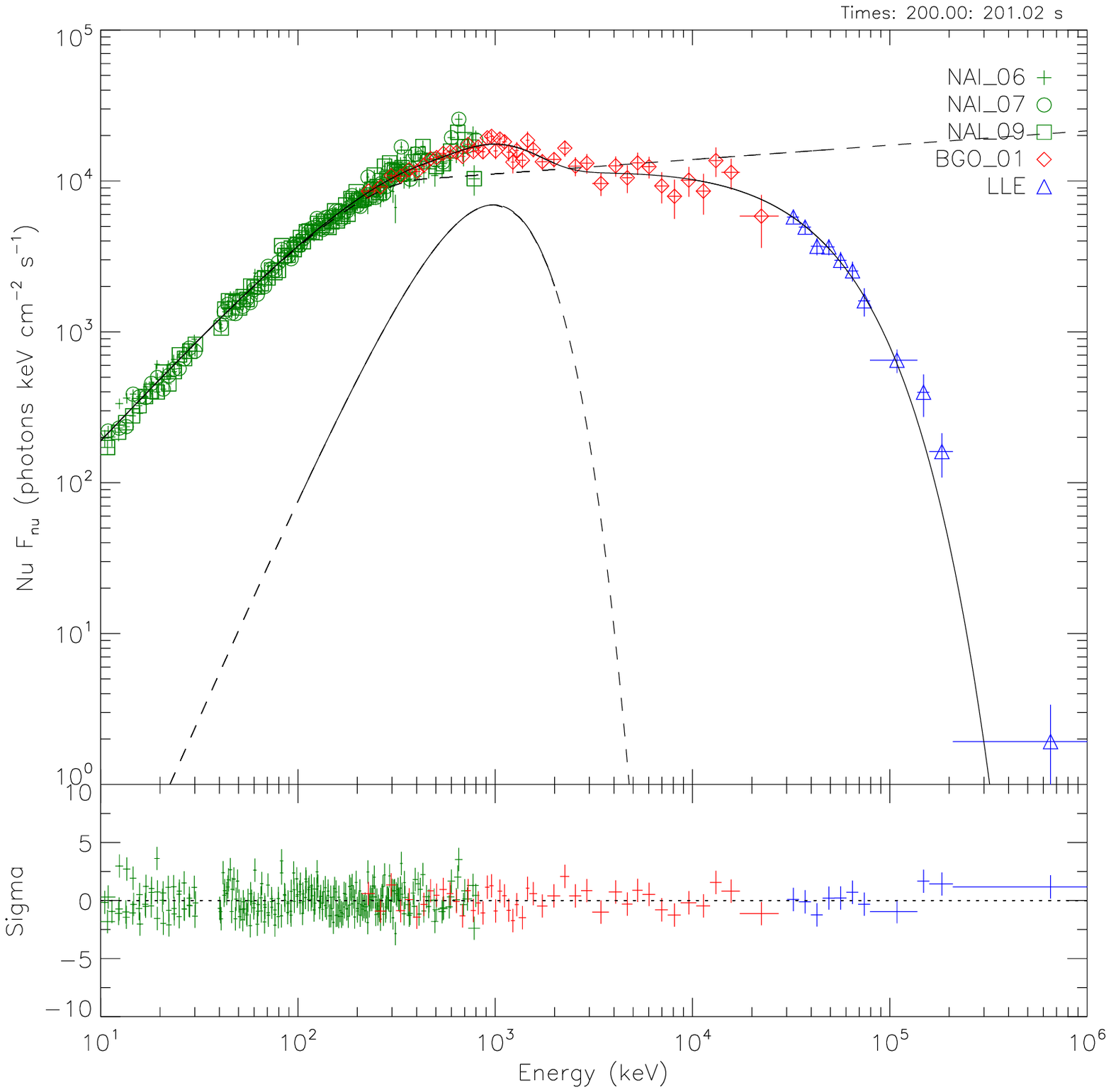}
\includegraphics[angle=0,scale=0.3]{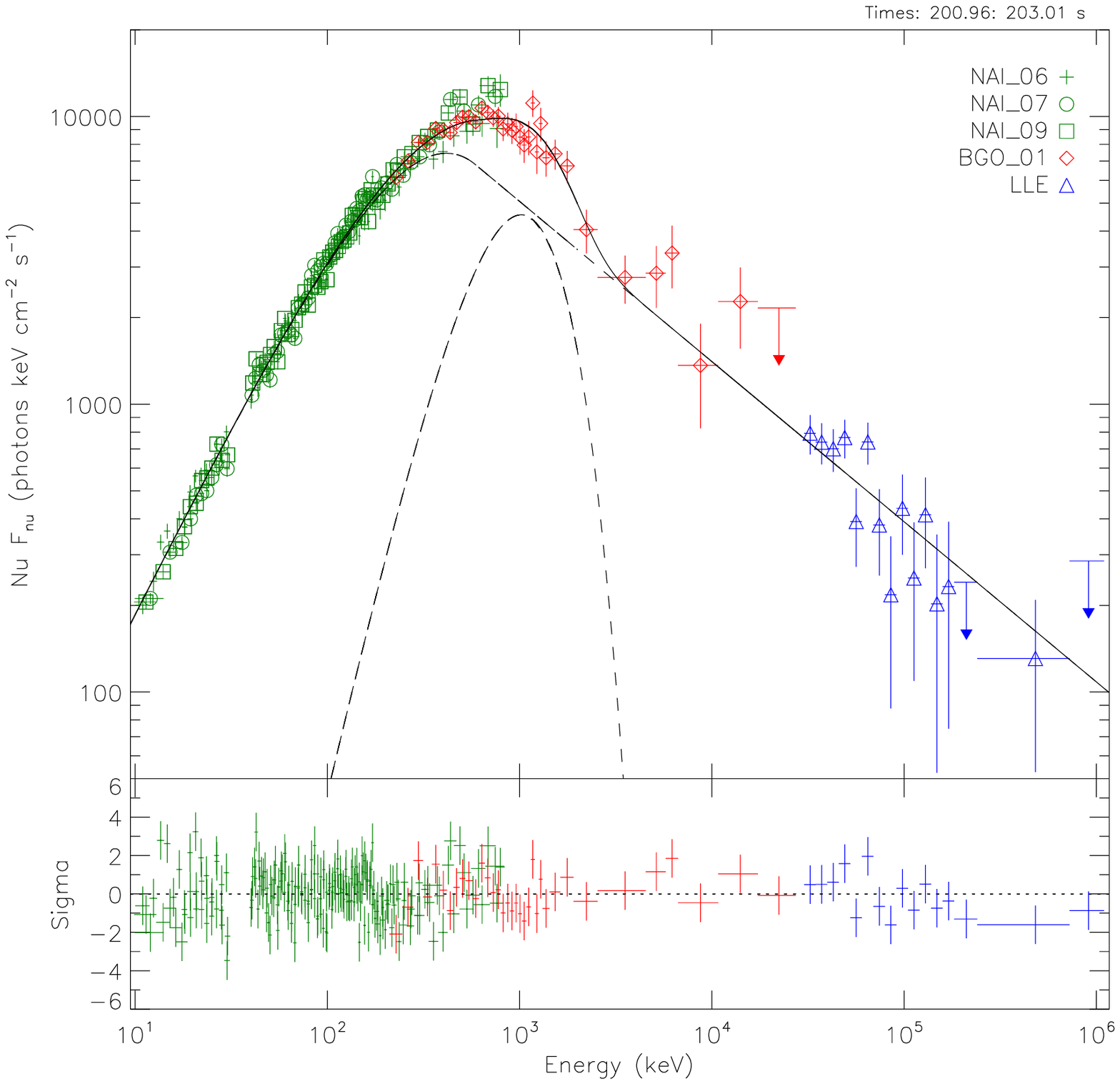}
\caption{Time-resolved spectra of GRB 160625B fitted with the assumed models in Table.1}
\hfill
\end{figure}

\clearpage

\begin{figure}[ht!]
\figurenum{2}
\centering
\includegraphics[angle=0,scale=0.48]{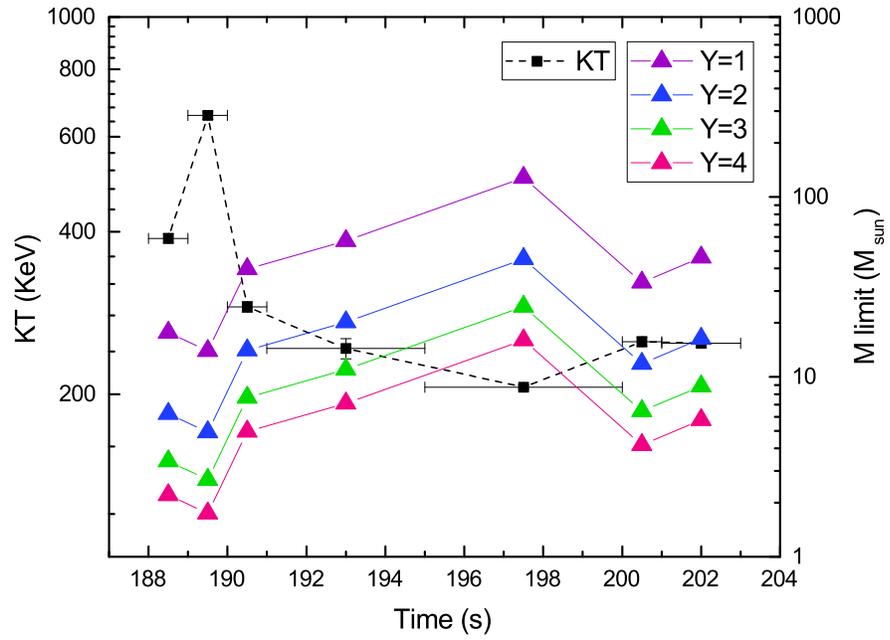}
\caption{The evolution of thermal emission temperature and the upper limit on the central black hole mass with the time.}
\hfill
\end{figure}

\begin{figure}[ht!]
\figurenum{3}
\centering
\includegraphics[angle=0,scale=0.48]{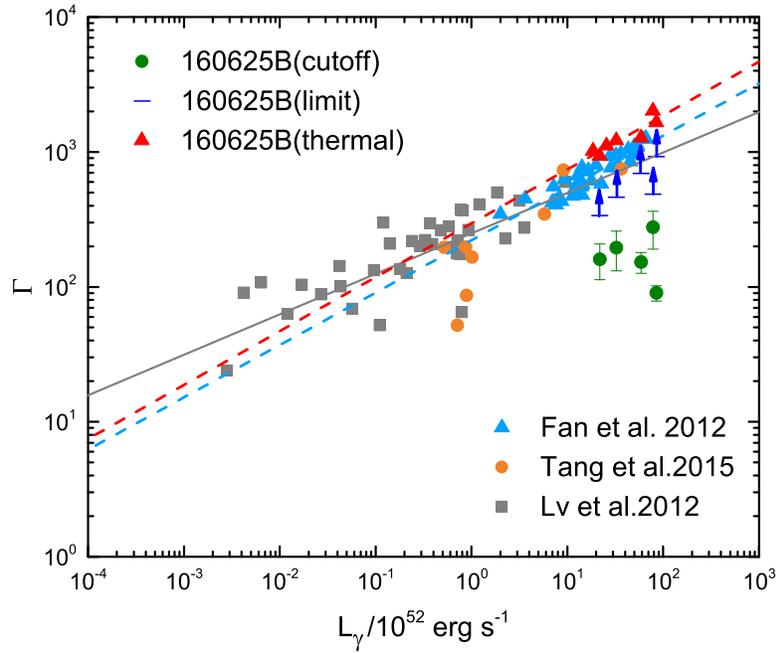}
\caption{The $\Gamma-L_\gamma$ correlation. Note that the Lorentz factors derived from the cutoffs are well below the lower-limits set by the electron pair Compoton scattering process. The Lorentz factors yielded in thermal-radiation modeling are well consistent with the $\Gamma-L_\gamma$ correlation holding for other bursts and in particular the distinct time-resolved thermal components identified in GRB 090902B.}
\hfill
\end{figure}

\begin{figure}[ht!]
\figurenum{4}
\centering
\includegraphics[angle=0,scale=0.48]{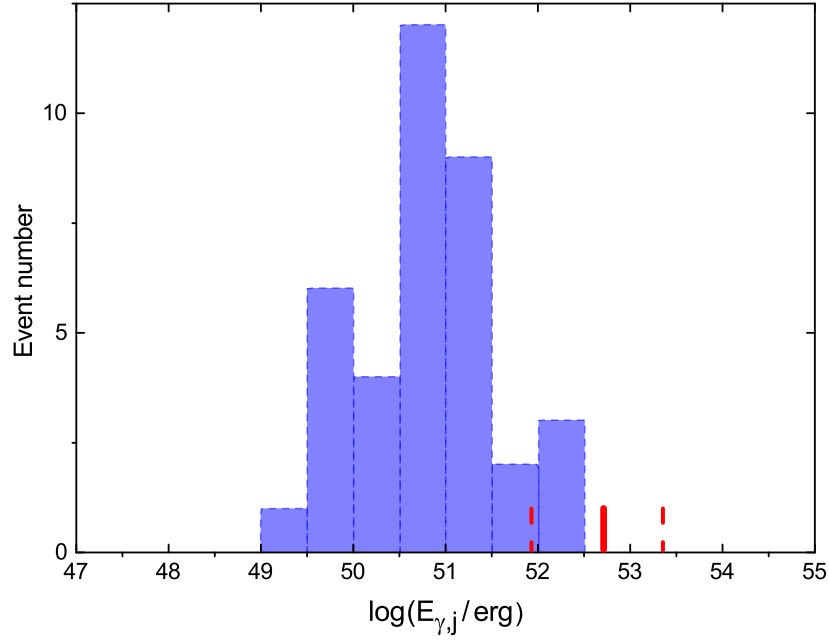}
\caption{The distribution of geometry-corrected gamma-ray energy (blue bars, estimates based on the sample presented in \citet{2016ApJ...818...18G}) comparing to the intrinsic prompt emission energy of GRB 160625B ($5.2\times 10^{52}$ erg, red solid line). Please note that with the half-opening angle found in the numerical fit of the late-time afterglow of GRB 160625B we have $E_{\gamma, \rm j}\sim 8\times 10^{52}$ erg.}
\hfill
\end{figure}

\begin{figure}[ht!]
\figurenum{5}
\centering
\includegraphics[angle=0,scale=0.48]{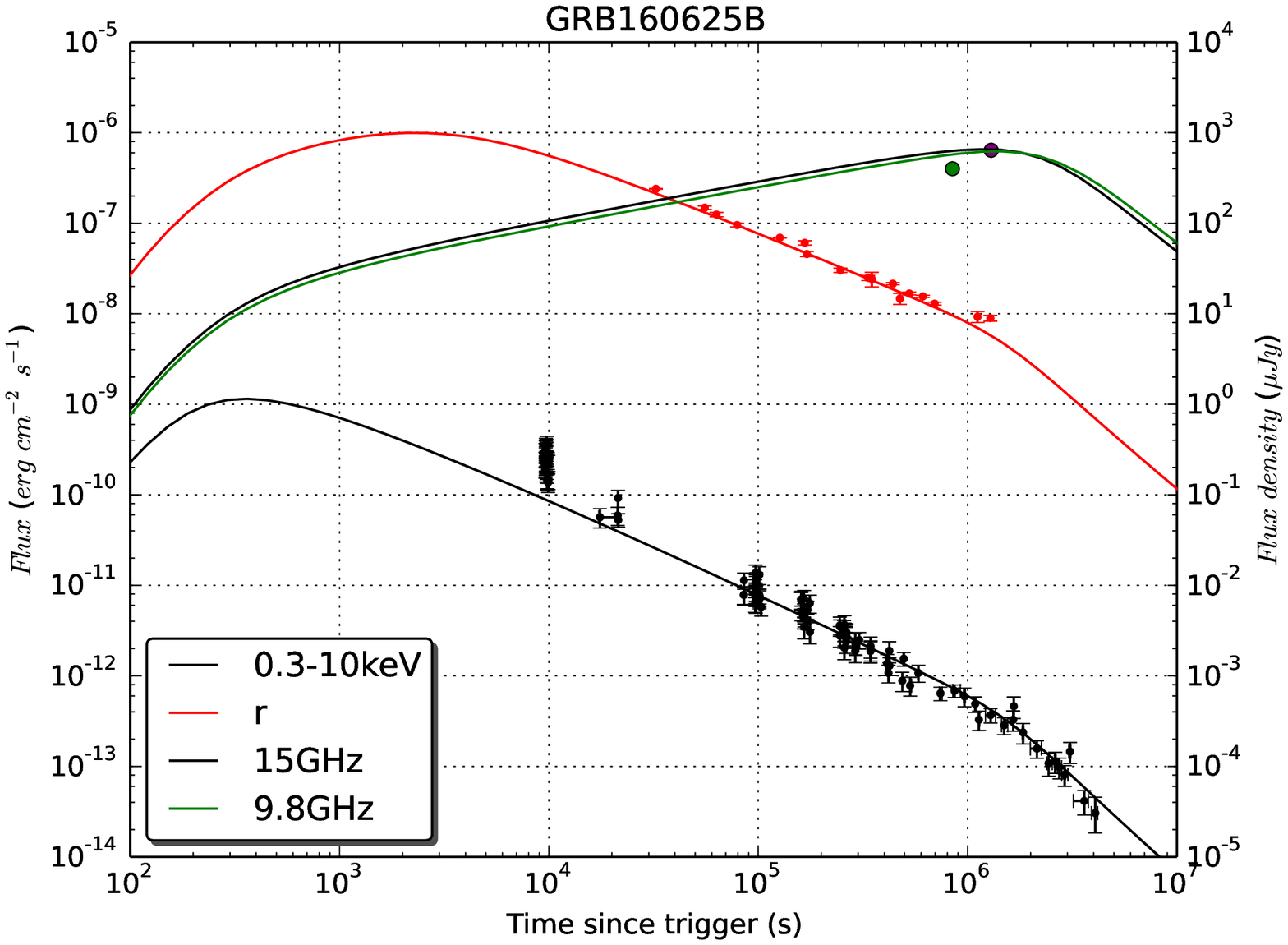}
\caption{Modeling the multi-wavelength afterglow of GRB 160626B with forward shock model. The black dots are X-ray(0.3-10 keV) data. The red dots are optical(r-band) data collected from GCN \citep{g1,g3,2016GCN...19602..1W,g7,g8, 2016GCN19619, 2016GCN19620, 2016GCN19640, 2016GCN19642, 2016GCN19651, 2016GCN19680}. The purple and green dots are radio afterglow data of 15GHz\citep{2016GCN...19610..1M} and 9.8GHz \citep{2016GCN19606} respectively.}
\hfill
\end{figure}

\end{document}